\documentstyle[aps,twocolumn]{revtex}
\begin{document}
\title{Disordered Electrons in a Strong Magnetic Field:
 Transfer Matrix  Approaches to the Statistics of the Local Density of States}

\author{Axel Dohmen$^1$, Peter Freche$^1$, Martin Janssen$^{1,2}$}

\address{$^1$ Institut f\"ur Theoretische Physik, Universit\"at zu
K\"oln, Z\"ulpicher Strasse 77, 50937 K\"oln, Germany \\\
$^2$ Department of Physics, Technion, 32000 Haifa, Israel}
\date{{25.1.96}}
\maketitle

\begin{abstract}
We present two novel approaches to establish the local density of
states as an order parameter field for the Anderson transition
problem.
We first demonstrate for 2D quantum Hall systems the validity of
conformal scaling relations which are characteristic of order
parameter fields. Second we show the equivalence between the
critical statistics of eigenvectors of the Hamiltonian and of the
transfer matrix, respectively.  Based on this equivalence we obtain
the 
order parameter exponent $\alpha_0\approx 3.4$ for 
3D quantum Hall systems.
\end{abstract}

\pacs{71.50; 73.40.H; 61.43.H}

The absence of diffusion in coherent
disordered electron systems is known
as Anderson localization \cite{And}.
In dimensions $d>2$ a disorder induced
localization-delocalization (LD) 
transition occurs quite generally at some
value of the Fermi energy \cite{Kra}.
In $d=2$ all states are localized
unless a certain amount of spin-orbit
 scattering \cite{FasSch}
or a strong magnetic field
is present \cite{HucR,JanB}.
 The LD transitions of independent 
(spinless) 2D electrons subject
to a strong perpendicular magnetic field
are located 
at the Landau energies.
These transitions are generally believed
to be responsible for the
integer quantum Hall effect 
\cite{Pra,JanB}. 

\medskip

In general, LD transitions are 
characterized by the critical exponent
 $\nu$ of the localization length 
\cite{Abr} and by the multifractal
$f(\alpha)$ spectrum of the local
amplitudes of critical eigenstates 
\cite{JanR,Gru}.
 The
$f(\alpha)$ spectrum describes the statistics and scaling behavior
of the local density of states (LDOS). 
Although the average density of states does not reflect
the LD transition, the {\it typical}
 value (i.e.~geometrically averaged) of the LDOS does: it vanishes with an
exponent  $\beta_{\rm typ}=(\alpha_0-d)\nu$
on approaching the LD transition point, where $\alpha_0>d$ is the 
maximum position of $f(\alpha)$.
It is thus tempting to interpret the LDOS as an order parameter field
of the LD problem \cite{JanR} (see also \cite{Weg,Mir}).
Our aim  is to support this interpretation 
by establishing two  
characteristic features of order parameter fields 
for the LDOS: First, scaling exponents of the order parameter field
are related to the critical exponents  of the corresponding
 spatial correlation 
functions. These correlation functions  show conformal
invariance. Second, the scaling exponents are  universal in the
sense of (one-parameter) scaling theory, i.e.~any local quantity
containing contributions from the relevant scaling field  shows
asymptotically the
same spectrum of scaling exponents.

\medskip

In this article  we demonstrate that
the $f(\alpha)$ spectrum of critical eigenstates
is related to  correlation functions
 in different geometries by
conformal invariance. We derive the conformal mapping relations
appropriate for a multifractal situation and check them numerically
in 2D quantum Hall systems (QHS).
Furthermore, we have calculated numerically $f(\alpha)$ for
the local components of transfer matrix eigenvectors in 2D QHS
and show that it coincides with $f(\alpha)$ of the Hamiltonian eigenstates.
Thus, these  two local quantities
share the same spectrum of scaling exponents although their
microscopic
construction is quite different. Our findings support
the identification of the  LDOS as an order parameter field
for the LD transition.  From $f(\alpha)$ calculations
for the transfer matrix eigenvectors in a 3D QHS we obtain the
characteristic order parameter exponent $\alpha_0\approx 3.4$.

\medskip

The multifractal analysis of the statistics of critical
eigenstates $\psi$ (see~e.g.~\cite{JanR}) usually
starts from considering the  box-probability, $P(L_b):=\int_{\rm
box}|\psi|^2$,  
of some box with linear size $L_b$. The box-probability 
 is normalized to the total volume $L^d$, $P(L)=1$.
At the LD transition the corresponding distribution function
$\pi(P;L_b/L)$  gives rise to power law scaling for the
moments  
\begin{equation}
\langle P^q(L_b)\rangle_L \propto (L_b/L)^{d + \tau(q)} \label{2}
\end{equation}
where $\tau(q)$ is non-linear. The distribution function 
can be described in terms of a
single-humped, positive, function $f(\alpha)$:
\begin{equation}
\pi(P;L_b/L)\, dP \propto (L_b/L)^{d-f(\alpha)} \, d\alpha \, ,
 \label{1}
\end{equation}
where $\alpha:=\ln P/\ln (L_b/L)$.
The multifractal spectrum
  $f(\alpha)$  is the Legendre transform of  $\tau(q)$.
Thus,  the statistics of critical eigenstates
is encoded in $f(\alpha)$ or equi\-valently in $\tau(q)$.
The  statistics of critical eigenstates
transforms to the statistics  of the 
LDOS  $\rho ({\bf r})=\Delta^{-1}|\psi({\bf r})|^{2}$
 at criticality. Here $\Delta$ denotes the mean level spacing at the
corresponding energy. It is a non-critical 
 quantity, $\Delta\propto L^{-d}$.
Consequently, the typical value, $\rho_{\rm typ}:=\exp\left\langle
\ln\rho\right\rangle \propto L^{d-\alpha_0}$  where
 $\alpha_0$ is the maximum position
of $f(\alpha)$.
Scaling $L$ with the localization length  shows that $\rho_{\rm typ}$ 
vanishes on approaching
the transition point 
with characteristic exponent $\beta_{\rm typ} =(\alpha_0-d)\nu$.
This observation motivates  to consider the LDOS as an order parameter
field of the LD transition \cite{JanR}.

\medskip

To make contact between $f(\alpha)$ and  correlation exponents 
we first make  the general scaling ansatz
for the  $q$-dependent spatial correlation in square-like geometries
\begin{equation}
\langle \rho^q({\bf r}) \rho^q({\bf r'}) \rangle_L
\propto a(r/L)\cdot r^{-{\tilde{z}(q)}} \, , r=|{\bf r}-{\bf r'}|\, .
\label{3}
\end{equation}
Here 
 $\tilde{z}(q)$ fulfills the scaling relation   \cite{JanR,PraC} 
\begin{equation}
\tilde{z}(q)=2(1-q)d+2\tau(q)\, .
\label{SCALR}
\end{equation}

In ordinary critical phenomena 
 the coefficient $a(r/L)$ will  approach a constant for 
$r \ll L \to \infty$. Then,
 standard conformal mapping arguments in 2D
\cite{Car} say that the corresponding correlator in a long strip of width $L_T$
with periodic boundary conditions in transverse direction decays like
\begin{equation}
 \exp \lbrack -2u/\xi^{[q]}(L_T)\rbrack \, ,
\label{4}
\end{equation}
where $u$ is the distance in longitudinal direction and $\xi^{[q]}(L_T)$
define generalized ($q$-dependent) localization lengths in quasi-1D.
The localization lengths are then related to $\tau(q)$ in the following way
\cite{JanR} (see also \cite{Lud})
\begin{equation}
\xi^{[q]}(L_T)/L_T = 2\lbrack \pi \tilde{z}(q)\rbrack^{-1}\, .
\label{5}
\end{equation}
Equation (\ref{5}) establishes a relation between generalized
localization lengths and $f(\alpha)$. 

In the following we will show
that this relation remains valid, although  the
coefficient
$a(r/L)$ in Eq.~(\ref{3}) cannot be treated as  a constant. 
In a multifractal situation 
the coefficient $a(r/L)$ in Eq.~(\ref{3}) remains sensitive to the
actual system size $L$ even for $L\to \infty$ (at the transition point).
It has been shown that
$a(r/L)\propto (r/L)^{y_2(q)}$ with $y_2(q)=d+\tau(2q)-2qd$ \cite{JanR,PraC}.
Thus, standard conformal mapping arguments do not apply.
This fact has been stressed recently by Zirnbauer \cite{Zir}.
We recall that the main idea behind conformal mapping arguments is an 
extension of a  homogeneity
law for correlation functions with respect to rescaling. Such a 
law exists also in  our case, since any rescaling of {\it all}
length scales in the correlator of Eq.~(\ref{3})
by the same scaling factor $s$ leads to
\begin{equation}
\label{6}
\langle \rho^q(s{\bf r}) \rho^q(s {\bf r'}) \rangle_{sL} = s^{-\tilde{z}(q)}
\langle \rho^q({\bf r}) \rho^q({\bf r'}) \rangle_{L} \, . 
\end{equation}
We  make now the plausible assumption
 that this law
 can be extended to conformal mappings
which are local scale transformations:  conformal mappings
transform the correlator of a geometry $\Omega$ 
to the corresponding correlator in geometry $\tilde{\Omega}$.
For large but finite 2D systems
we are led  to 
\begin{equation}
\frac{\langle \rho^q(w(z_1)) \rho^q(w(z_2)) \rangle_{\tilde{\Omega}}}
{\langle \rho^q(z_1) \rho^q(z_2) \rangle_{\Omega}} =
|w'(z_1)|^{-\frac{\tilde{z}(q)}{2}}|w'(z_2)|^{-\frac{\tilde{z}(q)}{2}}\, , 
\label{7}
\end{equation}
where $w(z)$ is any holomorphic function of complex coordinate $z$
and $w'(z)$ denotes the derivative.

By choosing $w(z)= (L_T/2\pi)\ln z$ which maps the plane onto
a strip straightforward calculations (cf.~\cite{Car})
show
that the correlator in the strip is given by
 expression~(\ref{4}) with an additional 
prefactor, $\exp\lbrack (2\pi y_2(q)/L_T)(u-L_s)\rbrack$,
 where $L_s$
is the strip length. Note, that the important relation Eq.~(\ref{5})
between $\xi^{[q]}$ and  $\tilde{z}(q)$ remains valid.

\medskip

To calculate the generalized localization lengths 
$\xi^{[q]}$ in strip-like (quasi-1D)  systems
($L_s$ being the length and $L_T\ll L_s$
being the width in the remaining $d-1$
directions)
one can
 start from a Hamiltonian $H$.
The Green's function $G({\bf r},{\bf r'};E):= \langle {\bf r'}| \left(
E-H+i0^{+}\right)^{-1}|{\bf r}\rangle$ can be calculated for 
${\bf r},{\bf r'}$ situated at opposite ends of  a strip 
by a recursive method \cite{MacK}. The resulting Green's function
$G(L_s;E)$ yields (cf.~expression (\ref{4}))
\begin{equation}
\left( \xi^{[q]}\right)^{-1} = -(2L_s)^{-1} \ln \langle |G(L_s)|^{2q}\rangle
=: \lambda^{[q]}
\label{90}\, .
\end{equation}
By Eq.~(\ref{5}),  $\lambda^{[q]}$ is proportional to $\tilde{z}(q)$
which becomes negative for $q > 1$. This   corresponds
to those  rare events where the Green's function increases 
with $L_s$. To detect a reasonable amount of these
rare events one has to restrict the numerical
calculation to strip-lengths $L_s$ of a few typical localization lengths.

Alternatively, one can construct a transfer matrix $T$ which relates
either
the values
of the 
wavefunction at opposite ends of the strip or relates the
corresponding scattering states \cite{PicS,Cha}.
For a transverse width $L_T=M l$, $l$ being a microscopic scale
of the problem, the dimension of the transfer matrix $T$ is
proportional
to $M$.
Lyapunov exponents, $\mu_{i}$, can be defined
from the eigenvalues of $TT^{\dagger}$ corresponding to
eigenvectors ${\bf u}_i$,
\begin{equation}
TT^{\dagger} {\bf u}_i = e^{-\mu_i} {\bf u}_i
\label{10}\, .
\end{equation}
The smallest positive Lyapunov exponent, denoted by $\mu$, determines
the generalized inverse localization lengths \cite{Vul},
\begin{equation}
\lambda^{[q]}= -(2L_s)^{-1} \ln \langle e^{-q\mu} \rangle \, .
\label{11}
\end{equation}

We have calculated $\lambda^{[q]}$  for two models of QHS.
The first model is the random Landau model (RLM) describing 
the Hamiltonian of 2D disordered electrons restricted to
the Hilbert space of one Landau band. The disorder is caused by
a white-noise potential. The characteristic microscopic scale is the
magnetic length $l_b$. A detailed
description of this model is given e.g.~in Ref.~\cite{HucR}.
We applied the recursive Green's function method (RLM/GF) and a transfer matrix
method (RLM/TM). The second model is the network model (NWM) of
Chalker and Coddington \cite{Cha} (with a characteristic
microscopic
scale $l$) which is formulated in terms
of a  transfer matrix.

In Fig.~1. we show the corresponding $\lambda^{[q]}  L_T$ curves  and its
first  derivative for the NWM in the regime  $|q|\leq 1$.
The conformal mapping relation, Eq.~(\ref{5}), allows for the
calculation of $f(\alpha)$ 
and, especially, of the most interesting scaling exponent $\alpha_0$.
For comparison  with related work \cite{PokHucSKlePra}
 we also give the values for
 $\tau(2)$. We find within the RLM by the GF method 
$\alpha_0=2.28 \pm 0.03$   for strip widths of $100 l_B$
 and by the TM method  $\alpha_0=2.3
\pm 0.03$ and $\tau(2)=1.66 \pm 0.03$
 for strip widths of $50 l_B$. For the NWM
we
obtain
  $\alpha_0=2.27 \pm 0.01$ and $\tau(2)=1.64 \pm 0.03$ for strip widths
of $64 l$.   
The results agree  for the different models and agree 
with the values reported in the literature \cite{PokHucSKlePra}.
We emphasize  that our results confirm  the
 conformal mapping relation, Eq.~(\ref{5}).

\medskip

In three dimensions  only a few conformal mappings exist. Especially,
 for three dimensional strips with periodic boundary conditions
we have no appropriate conformal mapping  at our disposal.
We saw that conformal invariance is  sufficient to derive 
relations between $f(\alpha)$ and $\lambda^{[q]}$,
 however it may not be necessary.
 Therefore, we found it worthwhile to study if a similar relation
as Eq.~(\ref{5}) could be valid for 3D strip-like systems.
We thus  make the ansatz
\begin{equation}
 L_T  \lambda^{[q]} (L_T) = { C \left( \tau(q)
-3(q-1)\right)}\, ,
\label{20}
\end{equation}
where $C$ is some unknown constant.
Since  $\tau(q)$
has to vanish at $q=1$ and is smaller than $d(q-1)$ for $q>1$ this
ansatz raises the 
   question if
the generalized localization lengths $\lambda^{[q]}$ vanish for $q=1$
and become negative for $q>1$,  for the 3D problem too. 
 To check for such behavior we 
took a 3D version of the network model, introduced recently by Chalker and 
Dohmen \cite{ChaD}. The 
3D network model consists of coupled layers of 2D networks and the
corresponding transfer matrix is described in detail in Ref.~\cite{ChaD}.
Our numerical findings at LD transition points of the model
are shown in Fig.~2. They indicate
that $\lambda^{[q]}$  is compatible with  our ansatz.
Note that, so far, we have established a relation between
$f(\alpha)$ and $\lambda^{[q]}$ only at criticality in 2D.
Our 3D results suggest that a similar relation might be
valid in 3D at criticality. To us, it seems rather
unlikely that similar relations hold true  far off
criticality.

\medskip

Our second approach to establish the LDOS as an order parameter
for the LD transition is
based on the assumption that any local quantity in a 
given disordered system can be viewed as being decomposed in terms of local
 scaling fields  in the sense of critical phenomena theory (e.g.~\cite{Dom}).
If one-parameter scaling holds true then, on approaching criticality,
only the  relevant scaling field has to be considered giving rise to a
universal distribution of local amplitudes.
 Any contribution of irrelevant scaling fields
would lead to deviations between the corresponding
$f(\alpha)$ spectra, but eventually die out
in the thermodynamic limit.

  The components of the
eigenvectors ${\bf u}_i$ corresponding to the transfer matrix, Eq.~(\ref{10}),
 in the network model \cite{Cha} have local meaning (the corresponding
scattering channels are defined in real space) and can be subject to
a conventional 
multifractal analysis. 
From this one obtains a one-dimensional multifractal spectrum.
Multiplication with $d$ yields the 
associated  $d$-dimensional 
spectrum.

When considering the components of transfer matrix eigenvectors
which belong to large Lyapunov exponents (small localization lengths)
deviations  from a universal scaling behavior can clearly be
observed in the 2D network model.
For the smaller Lyapunov exponents, however,
we obtain almost identical results (see Fig.~3).
Concentrating on the eigenvector
${\bf u}$ corresponding to the 
smallest Lyapunov exponent, we find $\alpha_0=2.3\pm 0.02$ and
$\tau(2)=1.59\pm 0.05$. For the 2D network model
 we have thus evidence
 that eigenvector components of the transfer matrix give rise to
 the same $f(\alpha)$ spectrum as the  eigenstates
of the Hamiltonian.

\medskip

We also calculated  $f(\alpha)$
of transfer matrix eigenvectors for 
 the 3D network model at  LD transition points.
It is worth mentioning that the 3D network model is anisotropic
and for most values of the parameter that fixes the coupling between
adjacent layers
 the typical quasi-1d 
localization lengths are {\it smaller} than the system width
 $L_T$ in one of the
transverse directions, but 
the basic condition for the applicability of the 
boxing method is that the localization length has to be larger than $L_T$.
Restricting to those values of the coupling where this condition
was fulfilled we find from
the statistics of the eigenvector components
corresponding to the smallest Lyapunov exponent:
$\alpha_0 = 3.4\pm 0.2$ and $\tau(2)= 2.3\pm 0.2$ \cite{Comm}.
We mention that these values of $\alpha_0$ and $\tau(2)$ fulfill
the inequality $2/\tau(2) < \nu
 < (\alpha_0-d)^{-1}$
proposed by Janssen \cite{JanR} relying on one-parameter
scaling arguments ($\nu \approx 1.35$ \cite{ChaD}).

\medskip

In summary,  we  have proposed a conformal scaling relation,
Eq.~(\ref{5}), between the $f(\alpha)$ spectrum
of the local density of states at criticality  and the corresponding
generalized Lyapunov exponents in strip-like geometries.
We have 
 checked its validity for  2D quantum Hall systems by numerical
calculations. We have presented numerical results which support
the expectation that similar relations are valid in 3D systems at
criticality.
Furthermore, our numerical calculations show that eigenvector
components
of the transfer matrix give rise to the same $f(\alpha)$ as that
of the Hamiltonian eigenstates. 
We like to stress the following aspects of our results:  
First, they show that the local density of states
has several features in common with order parameter fields in ordinary
critical phenomena. Second, the very existence  of relations between the
critical statistics of eigenstates (or LDOS) of the Hamiltonian,
the generalized Lyapunov exponents,  and the critical statistics of
eigenvectors
of the transfer matrix is new and far from obvious. Third,
such relations provide alternative ways
of computing the $f(\alpha)$ spectrum of critical eigenstates. As an
application of this we calculated the critical order parameter
exponent
for 3D quantum Hall systems.

We thank
 John Chalker for pointing out to us that a conformal mapping relation
between
Lyapunov exponents and multifractal spectra should exist and Bodo
Huckestein for useful discussions. This work has been
supported by the Sonderforschungsbereich
341
and by MINERVA.

\noindent
Figure 1: Generalized inverse localization lengths
$\lambda^{[q]}  L_T$ 
(A) and first  derivative with respect to $q$ (B)
for
different strip widths $L_T=Ml$ in the 2D network model ($l$ is the
corresponding 
microscopic length scale).

\bigskip

\noindent
Figure 2: Generalized inverse localization lengths
$\lambda^{[q]}  L_T$
(A) and  first   derivative with respect to $q$ (B) for
different cross-sections  $(L_T)^2=(Ml)^2$  in the 3D  network model ($l$ is
the
corresponding
microscopic length scale).

\bigskip

\noindent
Figure 3: The exponent $\alpha_0$ describing the statistics of
eigenvector components corresponding to
 the $i$-th Lyapunov exponent in the 2D network model.


\begin{references}
\bibitem{Comm} The present  uncertainty of $\alpha_0-3$ in 3D does not
allow
for a reliable determination of the constant $C$ in Eq.~(\ref{20}).
\end{references}
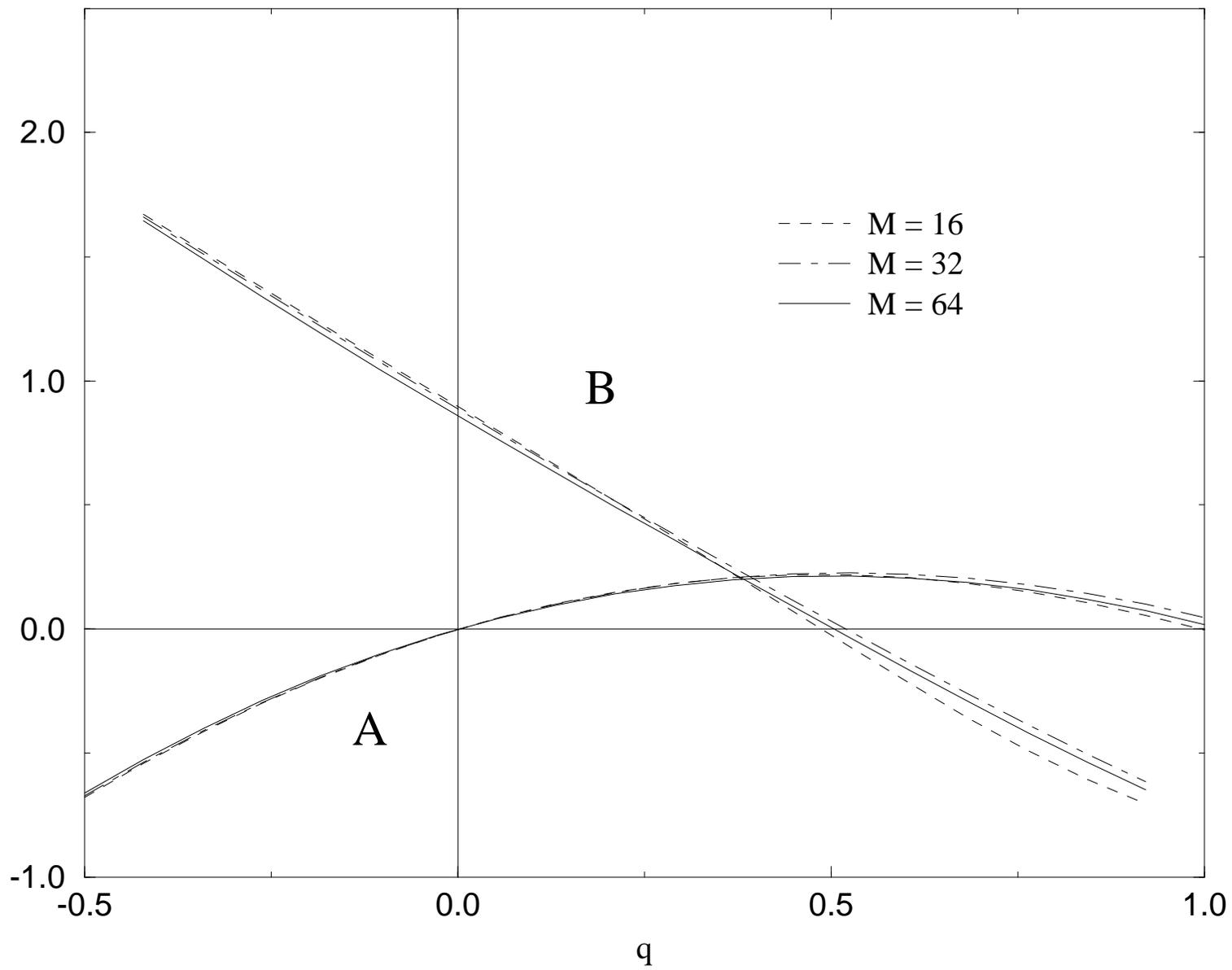
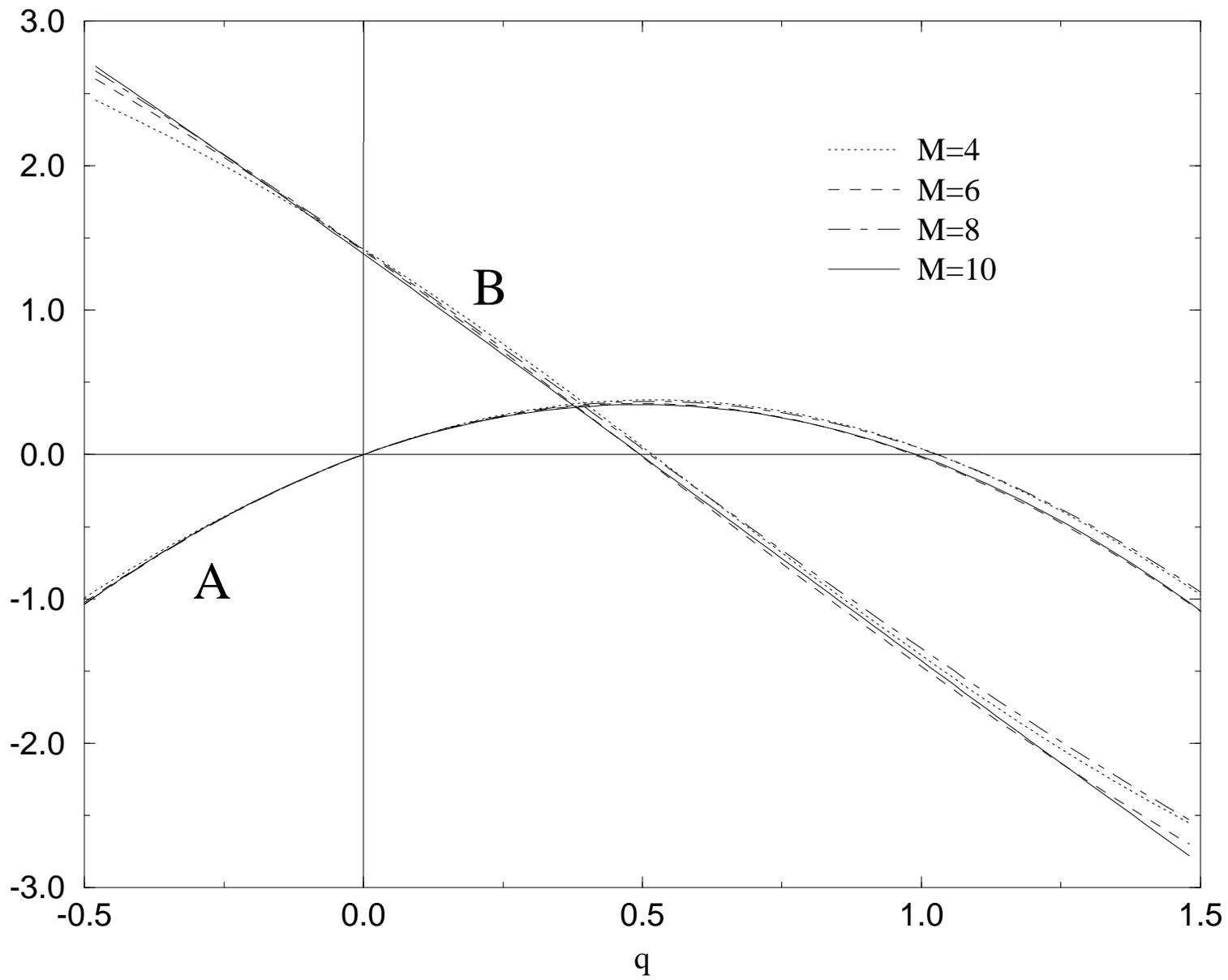
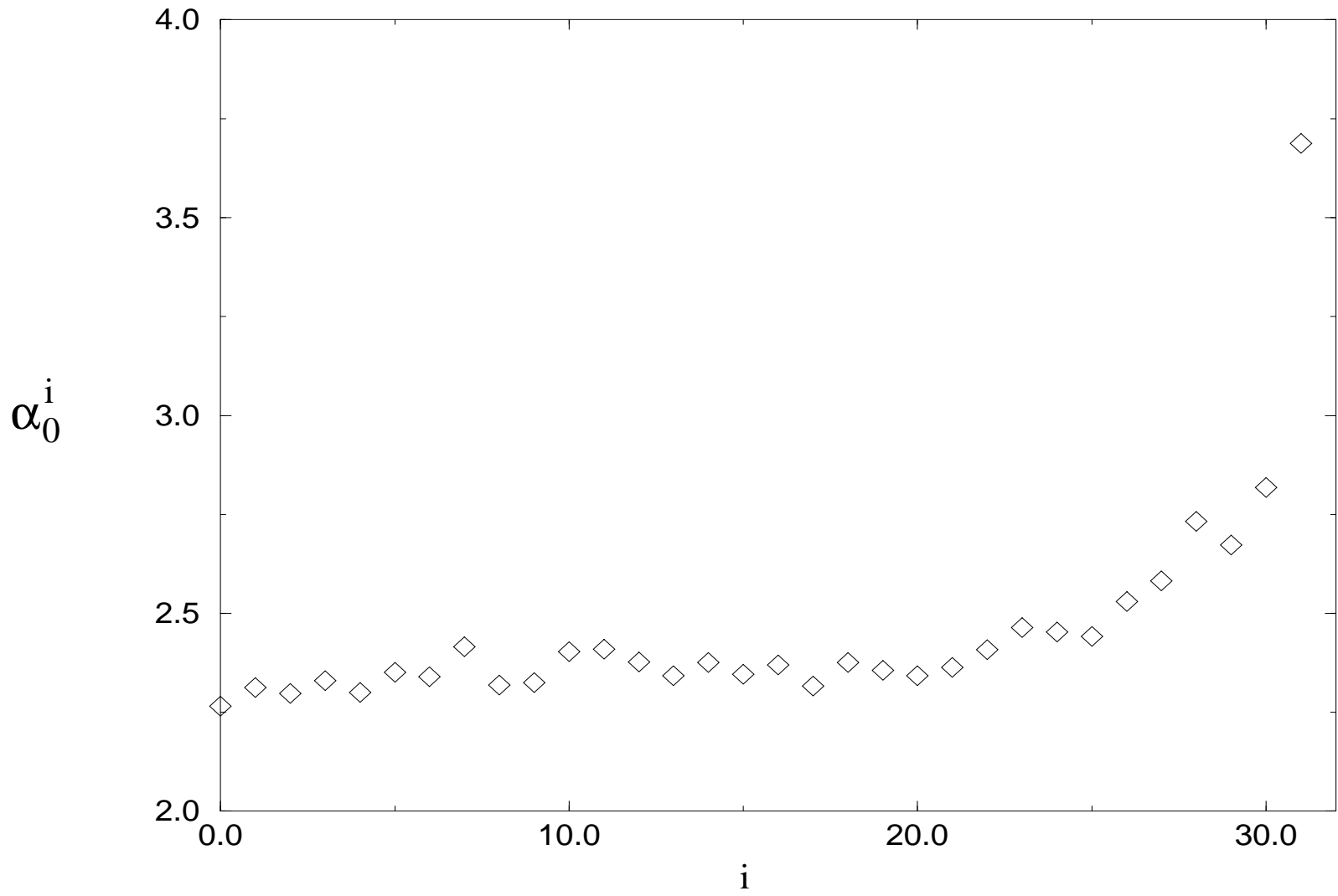
\end{document}